\documentstyle[12pt]{article}   

\large
\newtheorem{Theorem}{Theorem}[section]

\newtheorem{Lemma}{Lemma}[section]
\newcommand{\tiN}{\raisebox{-6.5pt}{$\displaystyle
\stackrel{\displaystyle N}{\sim}$}}

\newcommand{\beq}{\begin{equation}}
\newcommand{\eeq}{\end{equation}}

\makeatletter
\@addtoreset{equation}{section}
\makeatother

\title{Complete quantization of a diffeomorphism invariant field theory}  
\author{T. Thiemann\thanks{present address : Center for Gravitational Physics
and Geometry, The Pennsylvania State University, University Park,
PA 16802-6300, USA}
\thanks{thiemann@phys.psu.edu} \\
       Institute for Theoretical Physics, RWTH Aachen,\\
       52056 Aachen, Germany}
\date{{\small Preprint PITHA 93-33, August 93}}

\begin{document}

\maketitle                     

\begin{abstract}

In order to test the canonical quantization programme for general relativity
we introduce a reduced model for a real sector of complexified Ashtekar
gravity which captures important properties of the full theory. While it
does not correspond to a subset of Einstein's gravity it has the advantage 
that the programme of canonical quantization can be carried out
completely and explicitly, both, via the reduced phase space approach or 
along the lines
of the algebraic quantization programme.\\
This model stands in close correspondence to the frequently treated
cylindrically symmetric waves.\\
In contrast to other models that have been looked at up to now in terms of
the new variables the reduced phase space is infinite dimensional while the
scalar constraint is genuinely bilinear in the momenta.\\
The infinite number of Dirac observables can be expressed in compact and 
explicit form in terms of the original phase space variables.\\
They turn out, as expected, to be non-local and form naturally a set of
countable cardinality.

\end{abstract}

\section{Introduction}

Ashtekar's variables (\cite{1}) simplify the algebraic structure of the
constraint equations of general relativity so tremendously that one can
solve various problems of classical and quantum gravity that were simply
infeasible before in terms of the old ADM variables. In particular, a
number of model systems could be solved to more extent or even
completely in these new variables.\\
The model systems which could be solved completely up to now in terms
of the new variables (e.g. \cite{2}), to our knowledge, 
lack from not having in common at least one of the following features with 
general relativity :\\
1) the reduced phase space is infinite dimensional (i.e. the number of 
Dirac observables) and\\
2) the scalar constraint is bilinear in the momenta.\\
Here we introduce a new model which is an algebraic plus Killing reduction
of complexified Ashtekar gravity. In order to account for the fact that
Ashtekar's connection obeys a non-trivial reality condition, we also
impose reality conditions that bring us 'as close as possible' to the
cylindrically symmetric waves (\cite{3}) which form a genuine subset of
Einstein gravity whereas our model defines then a real sector of complexified
gravity which is manifestly no subset of Einstein gravity.\\
However, our model has both features 1) and 2) and we are thus able to
attack one of the major problems of general relativity, namely to find
a complete set of (preferredly canonically conjugate) Dirac observables for
a theory with an infinite number of degrees of freedom and non-trivial
dynamics. This is important in order to get some intuitive insights for the
problem of non-perturbatively quantizing full general relativity.\\
The results of this paper are as follows :\\

In section 2 we define the present model and show that it is not {\em quite}
possible to formulate the cylindrically symmetric waves (which means to
project our model of complex gravity to the real sector corresponding to
Einstein gravity) in terms of Ashtekar's variables after we carried out the
reduction. We then impose reality conditions that bring us at least
'as close as possible' to the cylindrically symmetric case. In particular,
the first- and the diagonal part of the second fundamental form are real
while the off-diagonal part of the latter is purely imaginary.

In section 3 we carry out the symplectic reduction (\cite{4}) of our model.\\
That is, we are using the degenerate Hamilton-Jacobi method to solve the 
constraints.\\
Quite surprisingly, it is straightforward to solve the typically very
complicated scalar constraint {\em before} solving the diffeomorphism
constraint. As usual for Hamiltonians or scalar constraints bilinear in
the momenta, there are several 'sectors' of the constraint surface. One of
them corresponds to degenerate metrics, the other to non-degenerate metrics
(the latter defines the 'physical' sector of our model).\\
The most complicated step is then to find a complete but not over complete
set of diffeomorphism invariant observables. This turns out to be feasible
and the resulting functionals of the basic phase space variables can
be given in explicit and compact form (the observables given in \cite{5}
are rather complicated and intractable objects due to the fact that
the mode system of Bessel functions is involved). \\
As expected for a diffeomorphism invariant field theory, the Dirac observables
come out to be naturally smeared and nonlocal. In contrast to non-diffeomorphism invariant field theories one does not have to employ a complete system of mode
functions on the hypersurface to write the Dirac observables in this form.
The reduced phase space thus has naturally a countable cardinality. We argue 
that this is always true for a spatially
diffeomorphism invariant field theory.

Section 4 is devoted to quantum theory.\\
From the reduced phase space point of view this is already trivial since the
phase space derived in section 3 can be cast into a form so that it has the 
structure of a cotangent space over a real, infinite dimensional vector 
space.\\
The application of the algebraic quantization programme forces us
to make use of functional integral techniques.\\
The two resulting Hilbert spaces turn out to be unitarily equivalent.\\
As to be expected for a field theory with constraints quadratic in the
momenta, the final set of Dirac observables mix configuration and momentum
variables with respect to the original polarization (here the Ashtekar
polarization).\\
While it is important to derive the Dirac observables, they are typically
hard to interpret. Here the framework of deparametrization (\cite{6})
helps to define 'evolving' Dirac observables which allow for a direct
interpretation in terms of metric variables and to leave the limits
set by the 'frozen' formalism.\\
We extend this formalism to field theories and apply it to our model. We
thus arrive at an interpretation of the theory although the meaning of the
constants of motion (Dirac observables) remains veiled.

Finally, in section 5, we discuss the relation to the loop representation 
(\cite{7}) of our
model.\\
First we define a set of loop variables that stand in bijection with the
phase space variables underlying the connection representation. This implies
that we can {\em explicitly} express constraints, symplectic form and
Dirac observables in terms of loop variables. These expressions are quite
complicated compared to those obtained for the connection representation and
that allows for two possible conclusions :\\
Either  
it seems to be doubtful whether the loop variables are tailored for the
problem to construct the reduced phase space or\\
Killing reduced models are unsuited to simulate the situation in the full
theory as far as the loop representation is concerned.

We summarize what has been learnt in dealing with the present model, valid
for the full theory of Einstein gravity.

\section{Definition of the model}

We assume the reader to be familiar with the Ashtekar formulation of
canonical gravity (\cite{1}). In the sequel a,b,c,.. will denote tensor
indices and i,j,k,.. SO(3) indices and we apply the abstract index
formalism.\\
We impose the following three conditions on the phase space variables
of the complexified Ashtekar formulation of gravity :\\
1) Killing-reduction : all fields are Lie-annihilated by a set of two
commuting, hypersurface orthogonal Killing fields.\\
Since the Killing fields commute they are surface forming. Since they are
hypersurface orthogonal these surfaces lie orthogonal to a foliating
vector field. Hence we find three globally independent coordinates x,y and z
such that the Killing fields are given by $\partial_x,\;\partial_y$.
Then all fields depend on z (and the time variable t) only.\\
2) algebraic reduction : We set to zero the following components of the
basic fields of the Ashtekar formulation
\beq A_z^1=A_z^2=A_x^3=A_y^3=E^z_1=E^z_2=E^x_3=E^y_3:=0 \;.\eeq
Up to now we are on the phase space of the so-called Gowdy models (\cite{8}).
We will assume, for simplicity, the spatial topology of the initial data
hypersurface to be that of the
three-dimensional torus $T^3$ in order to avoid technical difficulties 
associated
with boundary conditions.\\
One more condition brings us close to the axisymmetric spacetimes.\\
3) algebraic reduction : The Killing fields are orthogonal.\\
This last condition forces the densitized triad to be of the form
\beq
(E^a_i)=
\left ( \begin{array}{*{2}{c@{\mbox{ }}}c}
        E^x\cos(\beta) & -E^y\sin(\beta) & 0  \\
        E^x\sin(\beta) & E^y\cos(\beta) & 0 \\
        0 & 0 & E^z
        \end{array}
\right )
\eeq
and in order to have as many configuration variables as momentum variables we
require in analogy to (2.2) the Ashtekar connection to reduce to
\beq
(A_a^i)=
\left ( \begin{array}{*{2}{c@{\mbox{ }}}c}
        A_x\cos(\alpha) & -A_y\sin(\alpha) & 0  \\
        A_x\sin(\alpha) & A_y\cos(\alpha) & 0 \\
        0 & 0 & A_z
        \end{array}
\right )
\eeq
where all variables are generally complex up to now.\\
As far as the triads are concerned, provided we restrict them to be purely
real, we easily see that we have already a real diagonal 3-metric
(compare the appendix). Let us
see whether we can also implement the reality condition
\beq \mbox{Re}(A_a^i-\Gamma_a^i)=0,\;\Gamma_a^i=\mbox{ spin-connection} \eeq
which leads to the Einstein phase space. In order to do that we first have
to compute the spin-connection. The calculation is deferred to an appendix.
The result is given by
\beq
(\Gamma_a^i)=
\left ( \begin{array}{*{2}{c@{\mbox{ }}}c}
        -\Gamma_x\sin(\beta) & \Gamma_y\cos(\beta) & 0  \\
        \Gamma_x\cos(\beta) & \Gamma_y\sin(\beta) & 0 \\
        0 & 0 & \Gamma_z
        \end{array}
\right )
\eeq
where
\beq
\Gamma_x=\frac{1}{2 E^y}(\frac{E^y E^z}{E^x})',\;\Gamma_y=-\frac{1}{2 E^x}
(\frac{E^x E^z}{E^y})',\;\Gamma_z=-\beta'\;.
\eeq
The problem is already obvious : both, $\Gamma_a^i\mbox{ and }A_a^i$ have
{\em five} non vanishing components, giving rise to five reality conditions
induced by (2.4), but these components are coordinatized respectively by only
{\em four} independent coordinates (namely $A_x,A_y,A_z,\alpha$ and
$E^x,E^y,E^z,\beta$). The explicit form of these reality conditions is given
by the requirement that the following quantities ($iK_a^i:=iK_{ab}E^b_i/
\sqrt{\det(q)}=A_a^i-\Gamma_a^i\mbox{ where } K_{ab}$ is the extrinsic
curvature)
\begin{eqnarray}
& & iK_x^1:=A_x\cos(\alpha)+\Gamma_x\sin(\beta),
\;iK_x^2:=A_x\sin(\alpha)-\Gamma_x
\cos(\beta), \nonumber\\
& & iK_y^1:=-A_y\sin(\alpha)-\Gamma_y\cos(\beta),\;iK_y^2:=A_y\cos(\alpha)
-\Gamma_y\sin(\beta),\nonumber\\
& & iK_z:=A_z-\Gamma_z
\end{eqnarray}
are imaginary. Equivalently
\begin{eqnarray}
& &iK_x^i E^y_i=E^y(A_x\sin(\alpha-\beta)-\Gamma_x),\;iK_y^i E^x_i=E^x(-A_y
\sin(\alpha-\beta)-\Gamma_y),\nonumber\\
& & iK_x^i E^x_i=E^xA_x\cos(\alpha-\beta),\;iK_y^i E^y_i=E^y
A_y\cos(\alpha-\beta),\nonumber\\
& & iK_z^i E^z_i=E^z(A_z-\Gamma_z)
\end{eqnarray}
are imaginary.\\
We will now show that these are five functionally independent equations
whence our reduction is incompatible with (a subset of) the real form of
complexified gravity that corresponds to the Einstein phase space :\\
We will immediately show that the Gauss constraint of this model is given
by
\[ {\cal G}=(E^z)'-\sin(\alpha-\beta)(A_x E^x+A_y E^y)\;. \]
Let $\delta:=\alpha-\beta$. Then, using the Gauss constraint we derive
that
\beq \tan(\delta)=\frac{(E^z)'}{E^x[A_x\cos(\delta)]+E^y[A_y\cos(\delta)]}
\eeq
which is weakly imaginary since the square brackets in the denominator are
according to (2.8).
Assuming that the imaginary part of $\delta$ is finite, this implies
unambiguously that $\delta$ itself is imaginary :
$\delta=i\xi\mbox{ where }\xi$ is real. Then $\cos(\delta)=\mbox{ch}(\xi)$ is
real whence again from (2.8) we derive that $A_x, A_y$ are imaginary.
Finally we have $\gamma:=A_z+\alpha'=(A_z-\Gamma_z)+\delta'
=(A_z-\Gamma_z)+i\xi'$ which displays $\gamma\mbox{ and }A_z-\Gamma_z$ as
(weakly) imaginary quantities.\\
We still have to satisfy the two first conditions in (2.8). First of all,
again using the Gauss constraint, we find that
\begin{eqnarray}
& & \mbox{Re}(-A_x\sin(\delta)+\Gamma_x)=\frac{1}{E^x}\mbox{Re}(-A_x
E^x \sin(\delta)+E^x\Gamma_x)\nonumber\\
& = & \frac{1}{E^x}\mbox{Re}({\cal G}+A_y E^y\sin(\delta)-(E^z)'
+E^x\Gamma_x)\nonumber\\
& = & \frac{E^y}{E^x}\mbox{Re}(\frac{{\cal G}}{E^y}+A_y
\sin(\delta)+\Gamma_y) \nonumber\\
& \approx & \frac{E^y}{E^x}\mbox{Re}(A_y\sin(\delta)+\Gamma_y)
\end{eqnarray}
where we have used the identity $(E^z)'-\Gamma_x E^x+\Gamma_y E^y=0$ (which
is due to the covariant constance of the triads) and the reality of the
triads. This demonstrates that the real part of the first quantity in the 
first line of (2.10)
vanishes if and only if the last line in (2.10) is zero, at least on the
Gauss-constraint surface. However, since $A_x,A_y,\delta$ are imaginary,
the imaginary part of both expressions is already zero so that at least
one of the fields $A_x,A_y,\delta$ would be expressible completely in terms
of the triads and therefore the symplectic structure would become degenerate.
This furnishes the proof.\\
Since we are mainly interested in playing with a model that captures some
of the features of general relativity rather than describing the
cylindrically symmetric case,\footnote{The cylindrically symmetric case is
not of physical importance anyway because an axisymmetric matter distribution
is quite improbable} we take the following viewpoint :\\
We want to preserve at least the reality of the diagonal part of the
extrinsic curvature (the extrinsic curvature {\em is} diagonal for the
cylindrically symmetric waves). Since the metric is manifestly diagonal, we
can impose the simple reality condition that
\beq \alpha-\beta,\;A_z-\Gamma_z,\;A_x,\;A_y \eeq
are all purely imaginary.
Now, looking at (2.8), we see that
the extrinsic curvature will be symmetric upon imposition of the Gauss
constraint but it will develop an {\em imaginary} off-diagonal part
in the course of evolution. The only case when it is real is when
the off-diagonal part vanishes and we are then back on a genuine subset of
the phase space of the
cylindrically symmetric model (this subset is a fix point under evolution).
The diagonal part is real. The metric is diagonal
and real by construction so we managed to impose reality conditions that are
'as close' to the cylindrically symmetric case as possible.\\
These reality conditions are also preserved under evolution as we will
show shortly and that raises another problem :
Due to the evolution law of full general relativity
\beq \dot{q}_{ab}=2N K_{ab}+D_{(a}N_{b)} \eeq
an initially real, diagonal metric becomes complex and non diagonal when
inserting our $K_{ab}$ above. This is apparently a contradiction ! However,
there is a mistake in this argument : the processes of deriving the field
equations and reducing down the degrees of freedom cannot be expected to
commute if the phase space of the reduced model does not lie {\em entirely}
in that of the full theory as it is the case for our model.\\
One might ask whether it is possible to recover the cylindrically symmetric
case by making use of all SO(3) degrees of freedom (the three Euler angles
rather than only one, $\alpha\mbox{ and }\beta\mbox{ for } A_a^i\mbox{ and }
E^a_i$ respectively). However, if that was true
one could always go to the gauge (2.2), (2.3) and since the reality structure 
of Einstein gravity is SO(3) covariant (since SO(3) is a {\em real} Lie group)
one would obtain a contradiction.\\
Thus, it is not possible to treat the cylindrically symmetric case while
making use of a corresponding reduction on the Ashtekar phase space.\\
Nevertheless, this model has all features in common with full general
relativity (except for the modified reality conditions) so that it serves
as an interesting new testing ground for the quantization programme of
full quantum gravity as we will now show (actually this is also the viewpoint
that one takes in dealing with 2+1 gravity as a toy model for 3+1 gravity :
the connection of 2+1 gravity is purely real).

\section{Complete symplectic reduction of the model}

It will turn out that the symplectic reduction programme can be carried out
completely, that is, with respect to all constraints.\\
After integrating over the (finite) range of the coordinates x and y, the
reduced action becomes (\cite{1}; we neglect a trivial prefactor coming
from this integration)
\beq S=\int_R dt\int_{T^1}dz[-i\dot{A}_a^i E^a_i-(i\Lambda^i{\cal G}_i
-i N^a V_a+\tiN \frac{1}{2} C)] \eeq
where it is understood that we have to substitute for $A_a^i,E^a_i$ the
reduced expressions (2.2) and (2.3). Here ${\cal G}_i,\;V_a,\;C$ are,
respectively, the Gauss-, Vector- and scalar constraint and $\Lambda^i,
N^a,\tiN$ are the corresponding Lagrange multipliers, called respectively
the Gauss scalar potential, shift vector and $N:=\sqrt{\det(q)}\tiN$ the
lapse function.\\
By plugging the formulae (2.2) and (2.3) into the expressions for the
constraint functions we obtain
\begin{eqnarray}
{\cal G}_i & := & {\cal D}_a E^a_i:=\partial_a E^a_i+\epsilon_{ijk} A_a^j
E^a_k \nonumber\\
& = & \delta_{i3}[(E^z)'-\sin(\alpha-\beta)(A_x E^x+A_y E^y)]
                 =:\delta_{i3}{\cal G} \nonumber\\
V_a  &:=& F_{ab}^i E^b_i \nonumber\\
& = & \delta_a^z[\cos(\alpha-\beta)((A_x)'E^x+(A_y)'E^y)-A_z(E^z)'
         -\alpha'\sin(\alpha-\beta)((A_x)'E^x+(A_y)'E^y]\nonumber\\
    & = & \delta_a^z[\cos(\alpha-\beta)((A_x)'E^x+(A_y)'E^y)
          -\gamma(E^z)'+\alpha'{\cal G}]=:\delta_a^z V \nonumber\\
C & := & \epsilon_{ijk} F_{ab}^i E^a_j E^b_k \nonumber\\
& = & 2[A_x A_y E^x E^y+((A_x)'E^x+(A_y)'E^y)E^z\sin(\alpha-\beta)
\nonumber\\
& &   +\gamma(A_x E^x+A_y E^y)E^z\cos(\alpha-\beta)]=: 2C
\end{eqnarray}
where $F_{ab}^i$ is the curvature of the Ashtekar connection,
$\epsilon_{ijk}$ is the Levi-Civita symbol and we have abbreviated
$\gamma:=A_z+\alpha'$. Further we have absorbed the trivial factor of 2 in the
last line into the Lagrange multiplier of the scalar constraint.\\
Now we insert the reduced form of our basic coordinates into the symplectic
potential and obtain up to a total differential
(note that spatial integrations by parts do not contribute boundary terms
because the torus has no boundary)
\begin{eqnarray}
i\Theta[\partial_t] & = & \int_\Sigma dz[\dot{A}_z E^z
+\dot{A}_x E^x\cos(\alpha-\beta)\nonumber\\
& & +\dot{A}_y E^y\cos(\alpha-\beta)
-\dot{\alpha}(A_x E^x+A_y E^y)\sin(\alpha-\beta)] \nonumber\\
& = & \int_\Sigma dz[(\dot{A}_z+\dot{\alpha}') E^z
+\dot{A}_x E^x\cos(\alpha-\beta)+\dot{A}_y E^y\cos(\alpha-\beta)
+\dot{\alpha}{\cal G}] \nonumber\\
& =: & \int_\Sigma dz[\dot{\gamma}\Pi_\gamma
+\dot{A}_x \Pi^x+\dot{A}_y \Pi^y+\dot{\alpha}\Pi^\alpha] \;.
\end{eqnarray}
Note that all configuration variables on the Gauss-reduced phase space are
imaginary while the momenta are all real. This is a very nice reality
structure.\\
We next write the constraints in the so defined canonical
coordinates. For the Gauss and vector constraint this is easy :
\begin{eqnarray}
{\cal G} & = & \Pi^\alpha, \nonumber\\
V & = & (A_x)'\Pi^x+(A_y)'\Pi^y+\alpha'\Pi^\alpha-\gamma(\Pi^\gamma)'
\end{eqnarray}
while for the scalar constraint we have to do some more work.\\
We have directly from the definitions of our canonical coordinates
\beq \tan(\alpha-\beta)=\frac{(\Pi^\gamma)'-\Pi^\alpha}{A_x \Pi^x+A_y\Pi^y}
\eeq
so that we have managed to express the combination $\alpha-\beta$ in terms
of canonical variables.\\
Next, we use the trigonometrical identities
\[ \sin^2=\frac{\tan^2}{1+\tan^2},\;\cos^2=\frac{1}{1+\tan^2} \]
and have unambiguously
\begin{eqnarray}
C & = & A_x A_y\frac{\Pi^x\Pi^y}{\cos^2(\alpha-\beta)}+((A_x)'\Pi^x
+(A_y)'\Pi^y)\Pi^\gamma\tan(\alpha-\beta)+\gamma(A_x\Pi^x+A_y\Pi^y)\Pi^\gamma
\nonumber\\
& = & A_x A_y\Pi^x\Pi^y(1+[\frac{(\Pi^\gamma)'-\Pi^\alpha}{A_x \Pi^x
+A_y\Pi^y}]^2)+((A_x)'\Pi^x+(A_y)'\Pi^y)\Pi^\gamma \nonumber\\
& & \frac{(\Pi^\gamma)'-\Pi^\alpha}{A_x \Pi^x+A_y\Pi^y}
+\gamma(A_x\Pi^x+A_y\Pi^y)\Pi^\gamma\nonumber\\
& = & (\frac{1}{A_x \Pi^x+A_y\Pi^y})^2
A_x A_y\Pi^x\Pi^y([(\Pi^\gamma)'-\Pi^\alpha]^2+[A_x \Pi^x+A_y\Pi^y]^2)
\nonumber\\
& & +((A_x)'\Pi^x+(A_y)'\Pi^y)\Pi^\gamma[(\Pi^\gamma)'-\Pi^\alpha]
[A_x \Pi^x+A_y\Pi^y]+\gamma(A_x\Pi^x+A_y\Pi^y)^3 \Pi^\gamma\}\nonumber\\
& =: & (\frac{1}{A_x \Pi^x+A_y\Pi^y})^2 C \;.
\end{eqnarray}
We now rescale all configuration variables except for $\alpha$ by a factor of
i so that these become real too (e.g. $-iA_x$ will be called $A_x$ again.
Then the reduced action becomes
\beq S=\int_R dt\int_\Sigma dz[\dot{A}_x E^x+\dot{A}_y E^y+\dot{\gamma}
\Pi^\gamma+i\dot{\alpha}\Pi^\alpha-[i\Lambda\Pi^\alpha+N^x V-\tiN C]]
\eeq
with manifestly real constraint functions $\cal G$, V and C and Lagrange
multipliers.
This furnishes the proof alluded to in the previous section that the
reality structure of the model is preserved under evolution.\\
We will assume that the prefactor $1/(A_x \Pi^x+A_y\Pi^y)^2$
never diverges or vanishes so that we can absorb it into the Lagrange
multiplier and thus obtain a {\em fourth order} scalar constraint for this
model.\\
It is not trivial to check whether the constraint algebra closes. Before
doing that, we will pass to the following equivalent set of constraints,
obtained by subtracting a term proportional to the Gauss constraint
from the scalar and vector constraint :
\begin{eqnarray}
{\cal G} & = & \Pi^\alpha,\nonumber\\
V & = & (A_x)'\Pi^x+(A_y)'\Pi^y-\gamma(\Pi^\gamma)' \nonumber\\
C & = & A_x A_y\Pi^x\Pi^y(-((\Pi^\gamma)')^2+(A_x \Pi^x+A_y\Pi^y)^2)
+\gamma(A_x\Pi^x+A_y\Pi^y)^3 \Pi^\gamma\nonumber\\
& & -((A_x)'\Pi^x+(A_y)'\Pi^y)\Pi^\gamma(\Pi^\gamma)'
[A_x \Pi^x+A_y\Pi^y]
\end{eqnarray}
and something amazing happens when adding the term
$\Pi^\gamma(\Pi^\gamma)'[A_x \Pi^x+A_y\Pi^y]V$ proportional to the vector
constraint to the scalar constraint
\begin{eqnarray}
C & = & A_x A_y\Pi^x\Pi^y(-((\Pi^\gamma)')^2+(A_x \Pi^x+A_y\Pi^y)^2)
\nonumber\\
& & -\gamma\Pi^\gamma[(\Pi^\gamma)']^2
[A_x \Pi^x+A_y\Pi^y]+\gamma(A_x\Pi^x+A_y\Pi^y)^3\Pi^\gamma \nonumber\\
& = & [-((\Pi^\gamma)')^2+(A_x \Pi^x+A_y\Pi^y)^2]
 [A_x A_y\Pi^x\Pi^y+\gamma\Pi^\gamma(A_x \Pi^x+A_y\Pi^y)]
\end{eqnarray}
i.e. the scalar constraint factorizes into two second order constraints
with respect to the momenta ! They are given by
\begin{eqnarray}
C_1 & := & A_x A_y\Pi^x\Pi^y+\gamma A_x \Pi^x\Pi^\gamma+\gamma A_y\Pi^y
\Pi^\gamma\mbox{ and } \\
C_2 & := & -((\Pi^\gamma)')^2+(A_x \Pi^x+A_y\Pi^y)^2 \;.
\end{eqnarray}
By studying the transformations that these constraints generate, we obtain
that\\
$A_x,A_y,\gamma,\pi^x,\Pi^y,\Pi^\gamma,\Pi^\alpha$ are O(2)-invariant
while $\alpha$ is an O(2)-angle and that\\
 $A_x,A_y,\alpha,\Pi^\gamma$
are scalars while $E^x,E^y,\Pi^\alpha,\gamma$ are densities of weight one,
i.e. $\cal G$ generates O(2) transformations while $V$ generates
diffeomorphisms on the torus. By the way, $\gamma$ can
be interpreted as the abelian version of a connection plus the Maurer-Cartan
form $\theta_{MC}=+(dS)S^{-1}$ (which in one dimension for $S=\exp(\alpha)$
boils down to $\theta_{MC}=\alpha' dx$) and therefore is an O(2)-invariant
quantity.\\
The only nontrivial Poisson-bracket to check is then again that between
two scalar constraints (all constraints are O(2) invariant and
Diff($\Sigma$) covariant). We have
\begin{eqnarray}
& & \{C[M],C[N]\}=\int_\Sigma dx\int_\Sigma dy M(x)N(y)\{C_1(x)C_2(x),C_1(y)
C_2(y)\} \nonumber\\
& = & \int_\Sigma dx\int_\Sigma dy M(x)N(y)[C_1(x)C_1(y)\{C_2(x),C_2(y)\}
+C_1(x)C_2(y)\{C_2(x),C_1(y)\}\nonumber\\
& & +C_2(x)C_1(y)\{C_1(x),C_2(y)\}+C_2(x)C_2(y)\{C_1(x),C_1(y)\}] \nonumber\\
& = & \frac{1}{2}\int_\Sigma dx\int_\Sigma dy (M(x)N(y)-M(y)N(x))
[C_1(x)C_1(y)\{C_2(x),C_2(y)\}\nonumber \\
& & +2C_1(x)C_2(y)\{C_2(x),C_1(y)\}+C_2(x)C_2(y)\{C_1(x),C_1(y)\}] \nonumber\\
& = & \int_\Sigma dx\int_\Sigma dy (M(x)N(y)-M(y)N(x))
C_1(x)C_2(y)\nonumber\\
& & [-\delta_{,x}(x,y)]2(\Pi^\gamma)'(x)\Pi^\gamma(y)(A_x\Pi^x+A_y
\Pi^y)(y)\nonumber\\
& = & \int_\Sigma dx(M'N-MN')C_1C_2((\Pi^\gamma)^2)'(A_x\Pi^x+A_y\Pi^y)
\nonumber\\
& = & C[(M'N-MN')((\Pi^\gamma)^2)'(A_x\Pi^x+A_y\Pi^y)]
\end{eqnarray}
i.e. the constraints form a first class algebra, open in the BRST-sense, and
we can apply the framework of symplectic reduction (\cite{4}).\\
We begin with the Gauss-constraint :\\
We just have to pull back the symplectic potential to the $\Pi^\alpha=0$
surface, the vector and scalar constraint as they stand in (3.8) and
(3.9) are already reduced.\\
As experience with other model systems shows\footnote{specifically, the
author was looking at spherically symmetric gravity plus matter (\cite{9})},
it turns out to be much
more convenient to {\em first} reduce the scalar constraint and {\em then}
the vector constraint because the scalar constraint is not
diffeomorphism invariant but only diffeomorphism covariant.\\
We have to distinguish between two possible sectors : sector I is defined
by the vanishing of $C_1$, sector II by the vanishing of $C_2$.\\
We should mention that sector II corresponds to metrics that are degenerate
in the z-direction : For our model the metric tensor takes the form
\beq q_{ab}=\mbox{diag}(\frac{E^y E^z}{E^x},\frac{E^z E^x}{E^y},
\frac{E^x E^y}{E^z})\;..\eeq
From the Gauss-constraint and the definition of $\Pi_x,\Pi_y$ we infer
\beq
(A_x E^x+A_y E^y)^2=-[(E^z)']^2+[A_x\Pi^x+A_y\Pi^y]^2\mbox{ and }
\frac{E^x}{E^y}=\frac{\Pi^x}{\Pi^y} \eeq
so that we can solve $E^x,E^y,E^z$ in terms of reduced coordinates
\beq
(E^x,E^y,E^z)=(\pm\frac{\Pi^x}{A_x\Pi^x+A_y\Pi^y}\sqrt{C_2},
\pm\frac{\Pi^x}{A_x\Pi^x+A_y\Pi^y}\sqrt{C_2},\Pi^\gamma) \eeq
so that
\beq
q_{ab}=\mbox{diag}(\pm\frac{\Pi^y \Pi^\gamma}{\Pi^x},\pm\frac{\Pi^\gamma
\Pi^x}{\Pi^y},\pm\frac{\Pi^x\Pi^y}{[A_x\Pi^x+A_y\Pi^y]^2\Pi^\gamma}C_2) \;.
\eeq
Note that the quotient $\Pi^x/\Pi^y$ remains finite.\\
This shows that sector II is the 'unphysical' one.

\subsection{Sector I}

We pass to new canonical pairs,
\beq
(q_x:=\ln(|A_x|),p^x:=A_x\Pi^x;q_y:=\ln(|A_y|),p^y:=A_y\Pi^y;q:=\ln(|\gamma|),
p:=\gamma\Pi^\gamma)
\eeq
(note that $\delta\ln(|x|)=\delta\ln(x)\mbox{ for real x where }|x|:=
\sqrt{x^2}$) because then the scalar constraint $C_1$ adopts the simple form
\beq p^x p^y+p^y p+p p^x=0 \eeq
which does not even make any ordering problems any more if one would pass to the
quantum theory directly, that is using the Dirac approach. Moreover, it
defines trivially an abelian constraint because it consists only of momenta.
\\
Note that we could have changed the polarization for the pair $\gamma,
\Pi^\gamma$ in order that all configuration variables are scalars without
destroying the simple form of the scalar constraint, but then we would leave
the Ashtekar-polarization. Let us therefore proceed this way although this
will equip the canonical coordinates with a somewhat awkward tensor valence.
\\
It is clear that one will now diagonalize this constraint because then one
can solve the Hamilton-Jacobi equation by separation of variables. We have
\beq C=\frac{1}{4}[(p^x+p^y+2p)^2-(p^x-p^y)-(2p)^2] \eeq
so that we define still another set of canonical pairs
\beq P^0:=\frac{1}{2}(p^x+p^y+2p),\;P^1:=\frac{1}{2}(p^x-p^y),\;P^2:=p
\eeq
and conversely
\beq p^x=P^0+P^1-2P^2,\;p^y=P^0-P^1-2P^2,\;p:=P^2 \eeq
which allows for the determination of the new canonical configuration
variables via inserting (3.21) into the symplectic potential
\begin{eqnarray}
\Theta[\partial_t] & = & \int_\Sigma dx[\dot{q}_x(P^0+P^1-2P^2)+\dot{q}_y(P^0-P^1-2P^2)
+\dot{q}P^2]\nonumber\\
& = & \int_\Sigma dx[P^0\frac{d}{dt}(q_x+q_y)+P^1\frac{d}{dt}(q_x-q_y)
+P^2\frac{d}{dt}(q-2q_x-2q_y)]
\end{eqnarray}
from which we read off
\beq Q_0=q_x+q_y,\;Q_1=q_x-q_y,\;Q_2=q-2q_x-2q_y \; .\eeq
Then the scalar constraint just says that $(P^0,P^1,P^2)$ is a null-vector
in a 3-dimensional Minkowski space at every point of $\Sigma$. There are
two branches of general solutions to the associated Hamilton-Jacobi equation,
\beq
S_{\pm}[p^u,p^v;Q_\mu]=\int_\Sigma dz [\pm\sqrt{(p^u)^2+(p^v)^2}Q_0
+p^u Q_1+p^v Q_2]\eeq
where, following the second reference of \cite{4}, $p_u=P^1,p_v=P^2$ are
the new momenta and
\begin{eqnarray}
q_u & := & \frac{\delta S}{\delta p^u}=\pm\frac{p^u}{\sqrt{(p^u)^2+(p^v)^2}}
Q^0+Q^1\nonumber\\
q_v & := & \frac{\delta S}{\delta p^v}=\pm\frac{p^v}{\sqrt{(p^u)^2+(p^v)^2}}
Q^0+Q^2
\end{eqnarray}
are the new invariant (under the motions generated by the scalar constraint)
configuration variables.\\
One can get them also just by pulling back the symplectic structure
to the scalar constraint surface via its embedding $\iota$ into the phase
space (compare \cite{9}).
Inserting the solutions $P^0=\pm\sqrt{(P^1)^2+(P^2)^2}$ into the symplectic
potential yields (as before we do not display total differentials)
\begin{eqnarray}
& & (\iota^*\Theta)[\partial_t]=\int_{\partial_\Sigma}[\pm\dot{Q}_0
\sqrt{(p_u)^2+(p_v)^2}+\dot{Q}_1 p^u+\dot{Q}_2 p^v]\nonumber\\
& = & -\int_{\partial_\Sigma}[\dot{p}^u(\pm\frac{p^u}{\sqrt{(p_u)^2+(p_v)^2}}Q_0
+Q_1)+\dot{p}^v(\pm\frac{p^v}{\sqrt{(p_u)^2+(p_v)^2}}Q_0+Q_2) \nonumber\\
& = & \int_{\partial_\Sigma}[p^u\frac{d}{dt}(\pm\frac{p^u}{\sqrt{(p_u)^2
+(p_v)^2}}Q_0+Q_1)+p^v\frac{d}{dt}(\pm\frac{p^v}{\sqrt{(p_u)^2+(p_v)^2}}
Q_0+Q_2)\;.
\end{eqnarray}
It remains to reduce the vector constraint. Resubstituting all the
symplectomorphisms we have carried out up to now, we obtain
(note that for any two functions a and b holds
\begin{eqnarray*}
(\frac{a}{\sqrt{a^2+b^2}})'a+(\frac{b}{\sqrt{a^2+b^2}})'b
& = &  \frac{(a^2+b^2)'}{2\sqrt{a^2+b^2}})'-\frac{(a^2+b^2)'}
{2(\sqrt{a^2+b^2})^3}(a^2+b^2)=0\; )
\end{eqnarray*}
\begin{eqnarray}
V & = & (A_x)'\Pi^x+(A_y)'\Pi^y-\gamma (Pi^\gamma)' \nonumber\\
& = & (q_x)'p^x+(q_y)'p^y-\gamma (\frac{\Pi^\gamma}{\gamma})' \nonumber\\
& = & (q_x)'p^x+(q_y)'p^y+q'p-p' \nonumber\\
& = & (Q_0)'P^0+(Q_1)'P^1+(Q_2)'P^2-(P^2)' \nonumber\\
& = & \pm(Q_0)'\sqrt{(p^u)^2+(p^v)^2}+(Q_1)'p^u+(Q_2)'p^v-(p^v)' \nonumber\\
& = & (q_u)'p^u+(q_v)'p^v-(p^v)' \nonumber\\
& = & (q_u)'p^u+(\ln(|q_w|))'q_w p^w-(q_w p^w)' \nonumber\\
& = & (q_u)'p^u-q_w (p^w)'
\end{eqnarray}
where we have carried out one more canonical transformation
\beq q_v=:\ln(|q_w|),\;p^v=:q_w p^w \;.\eeq
Equation (3.28) displays $q_u,p^w$ as scalars and $q_w,p^u$ as densities of
weight one.

\subsubsection{Reality structure of sector I}

From their definition (3.20) it is clear that $p^x,p^y,p$ as products of
two real quantities are all real, whence also $P^\mu$ are real.\\
From (3.17) we have that
\beq Q_0=\ln(|A_x A_y|),\;Q_1=\ln(|\frac{A_x}{A_y}|),\;
Q_2=\ln(|\frac{\gamma}{(A_x A_y)^2}|)
\eeq
so also the $Q_\mu$ are real. Therefore, altogether, the left over canonical
pairs $(q_u, p^u; q_v, p^v)$ are both real and have range over the whole real
axis, whereas due to $q_w:=\exp(q_v)$ has only positive range while
$p^w:=p^v/q_v$ has still range over the whole real axis.

\subsection{Sector II}

Now the scalar constraint $C_2$ can even be written as one of two branches
of a constraint {\em linear} in momentum
\beq C_{\pm}:=A_x \Pi^x+A_y\Pi^y\pm (\Pi^\gamma)' \;. \eeq
Quite similar as for sector I we define new canonical pairs
\beq
(q_x:=\ln(|A_x|),p^x:=A_x\Pi^x;q_y:=\ln(|A_y|),p^y:=A_y\Pi^y;q:=\gamma,
p:=\Pi^\gamma) \eeq
and obtain
\beq C_{\pm}:=p^x+p^y\pm p' \;. \eeq
The general solution of the Hamilton-Jacobi equation is obviously given by
\beq S_{\pm}=\int_\Sigma dx[(p^v)'q_x+(p^w)'q_y\mp (p^v+p^w)q]
            =\int_\Sigma dx[p^v(\mp q-(q_x)')+p^w(\mp q-(q_y)')]
\eeq
whence we read off for the new momenta
\beq (p^v)'=p^x,\;(p^w)'=p^y \eeq
and for the new configuration variables
\beq q_v:=\mp q-(q_x)',\;q_w:=\mp q-(q_y)' \eeq
which are both densities of weight one so that the reduced (with respect to
the scalar constraint) vector constraint becomes
\beq V=-q_v(p^v)'-q_w(p^w)'=(q_u)' p^u-q_w(p^w)' \eeq
where we have defined still another canonical pair by
\beq q_v=:(q_u)',\;q^u:=-(p^v)'\;. \eeq
That this defines indeed a symplectomorphism is obvious from the following
short calculation (spatial surface integrals vanish due to our choice of
the spatial topology)
\beq \int_\Sigma \dot{q}_v p^v=\int_\Sigma (\dot{q}_v)'p^v
=\int_\Sigma \dot{q}_v(-p^v)' \; .\eeq

\subsubsection{Reality structure of sector II}

Again we have that $p^x,p^y,p$ are all real which then is true via
(3.34) also for $p^u,p^v$
and via (3.31), (3.35) also for $q_v,q_w$. Finally, via (3.37) also for 
$q_u,q_w$.
This time, however, $q_w$ is not necessarily positive.\\

\subsection{Reduction of the vector constraint}

We come now to the main problem for a spatially diffeomorphism invariant
field theory, namely to find a complete but minimal set of Dirac observables
which are to be promoted to the basic set of quantum operators later.\\
The stress of the following exposition is on the conceptual side. It is not
meant to be complete or rigorous to the last detail.\\
\\
It is always easy to find (classical) expressions that Poisson-commute
(even strongly) with the diffeomorphism constraint - the recipe is the
following : \\
1) From the transformation law of the basic field variables under
diffeomorphisms derive their tensor nature.\\
2) Construct scalar densities of weight one. \\
3) Integrate them over the initial data hypersurface.\\
These objects are then already invariant under diffeomorphisms that are
connected to the identity (in case of an asymptotically flat topology in
general only if the diffeomorphism tends to the identity rapidly enough at
spatial infinity). This is obvious from the transformation property of a
scalar density $\tilde{s}$ of weight one under an infinitesimal
diffeomorphism (generated by a vector field $\xi^a$)
$\delta\tilde{s}=\partial_a(\xi^a\tilde{s})$.\\
However, the scalar constraint is not at all so easy to solve since it
can be viewed as a dynamical constraint. Hence, it is most important to keep
the algebraic structure of the scalar constraint as simple as possible.\\
Now, since the scalar constraint is a density of weight two, it has a
nontrivial transformation law under diffeomorphisms. Accordingly, solving the
vector constraint before solving the scalar constraint will most likely
complicate rather than simplify the algebraic structure of the scalar
constraint because it will become a non-local object and, moreover,
cannot be reduced because it cannot be written in terms of diffeomorphism
invariant quantities.\\
The conclusion of all this is that the scalar
constraint is to be solved {\em before} the diffeomorphism constraint
for a field theory with a scalar constraint which is at least quadratic in
the momenta. This is what we were able to do up to now.\\
\\
Now we are going to derive a sequence of results which generalize in an
obvious manner (although this will become combinatorically much more
difficult) to higher dimensions.\\
\begin{Lemma}
The set of all integrated scalar densities constructed from the configuration
variables provides for a (possibly overcomplete) system of coordinates
of the diffeomorphism reduced configuration space.
\end{Lemma}
Proof :\\
1) Any local object constructed from the configuration variables has a
nontrivial transformation law with respect to infinitesimal diffeomorphisms.
Accordingly, diffeomorphism-invariant objects are non-local, i.e. they can
be expressed as integrals over (parts of) $\Sigma$ of local objects.\\
2) Were these integrals not over all of $\Sigma$ then they were not
invariant because the diffeomorphisms have their support everywhere except
for the boundary of $\Sigma$.\\
3) Only scalar densities of weight one are invariant when integrated over
$\Sigma$.\\
4) One could consider objects of the form
\beq \int_\Sigma dx_1 ...\int_\Sigma dx_m K(x_1,...,x_m) \eeq
i.e. $K(x_1,..,x_m)$ is some integral kernel. However, any such integral
kernel can be approximated arbitrarily well (in a given suitable norm by
an argument along the lines of the nuclear theorem) by
linear combinations of tensor products $f_1(x_1)..f_m(x_m)$ (for each k
the different functions $f_k$ will belong to a base). Thus, the integral of
K can be written as a sum of products of single integrals which displays K
as a derived object.\\
This proves the lemma.\\
$\Box$\\
Let $\{q^i\}_{i\in I}$ denote this set of Dirac observables
where the labelling is with respect to an index set I whose cardinality
is to be obtained. Let us further assume that all these observables are
algebraically independent. Any linear combination
\beq S:=\sum_{i\in I} p_i q^i \eeq
of these observables is again an observable where the $p_i$ are, in general,
complex numbers (the reality conditions on the $p_i$ match those on the $q^i$
so that there are as many independent momenta as configuration variables).
Then, provided that the set of the $q^i$ is complete in the sense
that every diffeomorphism invariant object constructed from the
configuration variables can be expressed by them, the above functional
S is the Hamilton-Jacobi functional relative to the diffeomorphism constraint
and the $p_i$ play the role of the integration momenta (recall the formalism
in the second paper of \cite{4}).\\
We will now define an overcomplete set of diffeomorphism invariant
functionals of the configuration variables for our model and then
systematically obtain a minimal subset. When stripping off the technical
methods that are particular for this model, one will get an insight for
what is necessary for the full theory.\\
According to lemma (3.1) what we have to do first is to construct all
possible scalar densities $\tilde{s}$ that can be built from $q_u,q_w$.
We first need two technical\\ lemmas :
\begin{Lemma}
1) In one dimension, tensors of valence (r,s) are scalar densities of weight
s-r.\\
2) In one dimension, a scalar density $\tilde{s}$ of weight one gives rise
to a $\tilde{s}$-compatible connection $\Gamma$ according to
\beq \Gamma:=[\ln(\tilde{s})]'\; . \eeq
\end{Lemma}
Proof :\\
1)\\
This lemma
is most easily proved by comparing the corresponding transformation
properties of tensors $T^{a..}\;_{b..}$ under diffeomorphisms
$x^a\rightarrow\tilde{x}^a=x^a-\xi^a$
\beq
\delta T^{a..}\;_{b..}={\cal L}_\xi T^{a..}\;_{b..}=\xi^c T^{a..}\;_{b..,c}
-T^{c..}\;_{b..}\xi^a_{,c}-..+T^{a..}\;_{c..}\xi^c_{,b}+..
\eeq
which in one dimension reduces to
\beq \delta T=\xi T'+(s-r)\xi' T \; .\eeq
On the other hand, a scalar density $\tilde{s}$ of weight n transforms as
\beq \delta\tilde{s}=\xi^a\tilde{s}_{,a}+n\tilde{s}\xi^a_{,a} \eeq
which again in one dimension reduces to
\beq \delta\tilde{s}=\xi\tilde{s}'+n\tilde{s}\xi' \eeq
and that furnishes the proof of the first part of the lemma.\\
Accordingly, in one dimension tensor valences are just characterized by
their density weight.\\
2)\\
One simply builds the unique
torsion-free connection compatible with a given metric by restricting the
Christoffel formula $\Gamma^a_{bc}=1/2q^{ad}(q_{db,c}+q_{dc,b}-q_{bc,d})$
to one dimension. A metric in one dimension is, according to 1), just
an arbitrary density of weight 2 corresponding to a symmetric tensor of
valence (0,2) (i.e. a metric $\tilde{\tilde{q}}$). Since the inverse of
a metric in one dimension is just
one over this metric we have for the affine connection in one dimension
$\Gamma=1/2(\tilde{\tilde{q}}'+\tilde{\tilde{q}}'-\tilde{\tilde{q}}')
/\tilde{\tilde{q}}=1/2[\ln(\tilde{\tilde{q}})]'$ or, by defining the density
$\tilde{s}$ of weight one through the equation
$\tilde{\tilde{q}}=\tilde{s}^2$, we have finally a connection defined by an
arbitrary density $\tilde{s}$ of weight one given by
\beq \Gamma:=[\ln(\tilde{s})]' \;.\eeq
Again, by just restricting the covariant derivative
\beq D_c T^{a..}\;_{b..}=\partial_c T^{a..}\;_{b..}+\Gamma^a_{cd}
T^{d..}\;_{b..}+..-\Gamma^d_{cb}T^{a..}\;_{d..}-..\eeq
to one dimension we find that for a density T of weight n we have
\beq DT=T'-n\Gamma T \eeq
and one can explicitly check that DT is a density of weight n+1, i.e. a
density again with one more covariant valence.\\
In particular $D\tilde{s}=\tilde{s}'-\Gamma\tilde{s}=0$ which shows that
the connection is $\tilde{s}$-compatible.\\
$\Box$\\
We are now ready to prove one of the main theorems of this section.
\begin{Theorem}
An (overcomplete) system of scalar densities of weight one is given by
\beq q_w f(\{(q_u)^{(n)}\}_{n\in{\cal N}})
\eeq
where f is an arbitrary function of the infinite number of arguments
indicated, $\cal N$ is the set of non-negative integers and for any scalar
function s, $s^{(n)}$ denotes the scalar $d^n(s)$ where d is the operator
$d=1/q_w\partial$ which is obviously anti-self-adjoint with respect to
the measure $d\mu=q_w dz$.
\end{Theorem}
Proof :\\
It is clear that any scalar density $\tilde{s}$ of weight one can be written
as $q_w s$ where s is a scalar, simply define $s:=\tilde{s}/q_w$. The
proof thus boils down to showing that the above function f in (3.49) is the
most general scalar that we can build.\\
Such a scalar will be the most general scalar constructed from $q_w,q_u$
and all their partial derivatives up to arbitrarily high order.
This can be done in a covariant way only by contraction of {\em tensors},
that is in one dimension, by multiplying densities such that their overall
weight is zero, according to lemma (3.2), first part. Hence we need
to construct a covariant derivative such that the derived tensors, i.e.
densities, transform as tensors, i.e. densities again. This can be done
by constructing an affine connection in one dimension and here part 2)
of lemma (3.2) comes into play.\\
We are now able to take arbitrarily high covariant derivatives of
$q_w,q_u$, for example with respect to the connection derived from
$\tilde{s}:=q_w$. But then $D q_w=0$ because D is $q_w$ compatible
which shows that the covariant derivatives of $q_w$ do not contribute
in our construction and furthermore $D^n\phi=q_w^n(1/q_w D)^n\phi=
q_w^n d^n\phi$ where we have introduced the operator $d=1/q_w\partial$.
\\
It remains to show that there is no loss in generality in choosing
$\tilde{s}:=q_w$.\\
Assume that we had chosen any other
density $\tilde{s}$ instead of $q_w$ in the construction of $\Gamma$
then the covariant derivatives $\tilde{D}$ of $q_w$ would not vanish, however
then $\tilde{D}q_w=\tilde{s}\partial(q_w/\tilde{s})
=\tilde{s}D(q_w/\tilde{s})=-q_w/\tilde{s}D\tilde{s}$ and $\tilde{s}$
is again built only from our configuration variables and D (it must be a
tensor and it cannot be constructed by using $\tilde{D}$ because $\tilde{D}$
is {\em defined} through $\tilde{s}$). Hence
$\tilde{D}q_w$ can be recast in the form of a function f of the type given
in (3.49). \\
This furnishes the proof.\\
$\Box$\\
What we have achieved up to now is to obtain a sufficient number of
observables $q^i$ which are functionals of the configuration variables
only.\\
We now ask for possible redundancies among them. Two sources of redundancy
are obvious :\\
1) by doing integrations by parts it is possible to relate some
of the $q^i$ and\\
2) there might be topological identities which could arise, for example, via
index theorems and thus affect the range of $q^i$ (for example in 2
dimensions the integrated densitized curvature scalar is an object of the
form we are considering here, but the Gauss-Bonnet theorem states that it
depends only on the genus of the compact hypersurface without boundary).\\
Under the assumption that we can restrict to analytical f's in (3.49)
it is possible to get rid of the first redundancy. What we are presupposing
is that there exists an expansion of f of the type
\beq f(\{a_m\}_{m\in{\cal N}})=\sum_{\{\{n_k\}\}}
f_{\{n_k\}}\prod_{k=0}^\infty a_k^{n_k}  \eeq
where $a_k:=d^k q_u$, the sum is over all sequences of (non-negative)
occupation numbers $n_k$ and the coefficients of the monomials
\beq M_{\{n_k\}}:=\prod_{k=0}^\infty a_k^{n_k} \eeq
are real numbers because $q_u,q_w$ are real as shown above.
Hence, in our case the index set is given by the set of sequences,
$I=\{\{n_k\}\}\ni i$, and the (still overcomplete) set of
$q^i$'s is given by the integrals
\beq q^i:=\int_\Sigma dz q_w M^i \eeq whereas the $f_i$ are the conjugate
momenta.\\
We will divide out the first, combinatorial redundancy by bringing the $q^i$
into a standard
form thereby introducing a terminology that one is used to from Fock-space
or statistical mechanics. :\\
First we observe that by doing an integration by parts the {\em occupation
numbers} $n_k$ of the {\em quantum states or energy levels}
k for {\em particles} of type $q_u$ may change
but that the {\em total particle number} (of each particle type separately
if there are more than 2 fields on the configuration space which it is not the
case in our situation)
\beq N:=\sum_{k=0}^\infty n_k \eeq
as well as the {\em total energy}
\beq E:=\sum_{k=0}^\infty k n_k \eeq
are preserved under such a process, simply because the number of factors
of a (irrespective of how often they are derived) and the number of
derivations does not change. In fact, one could write the operation
of doing an integration by parts in terms of {\em annihilators and creators}.
\\
\begin{Lemma}
Consider now the class of $(q^i)$'s with the same total particle number N and
total energy E. Then an algebraically independent genuine subset of
$q^i$'s in this class is given by those with indices i such that
$m_0=1..N,m_1=0$, the rest of the occupation numbers being arbitrary
(subject to the condition that they lie in the class labelled by (N,E).
\end{Lemma}
Proof :\\
In the first step we show that every member in the
given class can be written as a linear combination of these $q^i$'s and in
a second step we show that these $q^i$ cannot be transformed into each other.
\\
Consider the class $(N,E)=(M,k)$ and we may exclude the trivial case $M=0$.
Then $q^i$ is of the form
\beq q^i=\int d\mu\; a_{k_1}..a_{E-(k_1+..+k_{M-1})}
\eeq
where we do not care about occupation numbers, i.e. some of the
$k_r$ may be equal. Let, without loss of generality, $k_1$
be the lowest of the $k_r$. If $k_1=0$ we are done, otherwise do an
integration by parts. After that, $q^i$ is a linear combination (with integer
coefficients) of $q^j$'s of the required form since all these $q^j$ have
$m_0=1$. By doing further integrations by parts if necessary for these
$q^j$ (since $k (a_0)^{k-1} a_1=d((a_0)^k)$ can be integrated by parts)
one can satisfy $m_1=0, m_0>0$. This finishes the first step.\\
Let now one of these 'reduced' $q^i$'s be given. Then the set of its
occupation numbers is characterizing, i.e. by doing an integration by parts,
starting from $m_1=0$,
one necessarily picks up terms that have $m_1\not=0$ from the differentiation
of a since we already have $m_0>0$. This furnishes the proof.\\
$\Box$\\
From now on we will understand the index set I to be reduced, that is, we
consider the 'Mandelstam-identities' between the original $q^i$'s
originating from integrations by parts to be taken care of by lemma 3.3 .\\
As far as the second source of redundancy is concerned, we do not have any
indication that any of the $q^i$ should have a non-continuous range.\\
There is still a more subtle source of redundancy :\\
One could fix a gauge
(a frame) and expand $q_u$ into a taylor series with respect to this
frame where we have assumed these variables to be smooth functions on
$\Sigma$. The freedom in choosing these functions is then captured in the
choice of the values of their taylor coefficients.\\
However, then one could just integrate out all of our reduced
$q^i$ with respect to this frame and we would get complicated rational
functions
of these Taylor coefficients multiplied by real numbers coming from the
integration of the powers of the frame-variable x over the hypersurface
and depending only on the topology of $\Sigma$.\\
Since any such rational function can be generated from the Taylor
coefficients itself, the knowledge of the latter is sufficient and provides
for the minimal subset of observables.\\
The result of all this analysis summarized in the subsequent theorem.
\begin{Theorem} Only the following $q^i$'s are independent variables on the
diffeomorphism-reduced configuration space (n is any non-negative
integer), possibly modulo some discrete degrees of freedom :
\beq
q_n := \frac{1}{2}\int_\Sigma d\mu q_u d^{2n} q_u
=(-1)^n\frac{1}{2}\int_\Sigma d\mu [d^n q_u]^2\;.
\eeq
\end{Theorem}
Proof :\\
There is a one to one mapping between these functionals and
the Taylor coefficients of $q_u$, possibly up to sign. For example,
after fixing a gauge, e.g. $q_w(x)=g'(x)=1,\;
d=\partial$ where g is a fixed scalar function, the $q_n$ allow to extract
all the
Taylor coefficients of $q_u$ (e.g. $\partial^n\phi$ starts with the n-th
coefficient), possibly up to sign (that is, we might miss some discrete 
degrees of freedom).\\
$\Box$\\
Note that (3.56) are $|{\cal N}|$ {\em independent} quantities ($\cal N=$
natural numbers),
where $|S|$ denotes the cardinality of the set S, because they satisfy all
the criteria of the previous theorem to be algebraically independent and
precisely for that reason we have $b_{n/2}=\Phi_{n/2}=0$. Accordingly,
the number of degrees of freedom is {\em naturally countable} without 
employing a complete system of mode functions on $\Sigma$ to express the 
true degrees of freedom in discrete form.\\
Moreover, these observables have the nice feature that
\begin{eqnarray}
& & \frac{\delta q_n}{\delta q_u(x)}=q_w d^{2n} q_u(x), \\
& &  \frac{\delta q_n}{\delta q_w(x)}=-\frac{1}{2}\sum_{k=1}^{2n-1}(-1)^k
(d^k q_u(x))(d^{2n-1}q_u(x))\;.
\end{eqnarray}
We will denote the momenta conjugate to $q_n\mbox{ by } p^n$.\\
The associated Hamilton-Jacobi functional is
\beq S:=\sum_{n=0}^\infty p^n q_n \eeq
and the old momenta are the functional derivatives of S with respect to the
old coordinates.\\
\\
For sector I we have seen that $q_w$ is positive (and therefore serves as a
genuine density for a measure on $\Sigma$ in the the definition of $q_n$)
whereas for sector II
the range of $q_w$ is the whole real line. Since $q_u,q_w$ are both real,
$(d^n q_u)^2$ is positive whence $q_n$ has half or the full range of
the real line for sector I or II respectively. $p^n$ on the other hand is
generally real so that the pair $(q_n,p^n)$ coordinatizes the cotangent space
over the (positive) real line. We can avoid this complication for sector I
by going to the pair $(q_n,p^n)\rightarrow(\ln(|q_n|),q_n p_n)$ which we will
assume in the sequel.\\
\\
\\
Our, admittedly somewhat sketchy construction generalizes in an
obvious way to more
dimensions for every diffeomorphism invariant field
theory.

\section{Quantum theory and determination of the physical
inner product}

\subsection{Reduced phase space approach}

Quantum theory is straightforward. Keeping the polarization that we
have ended up with, physical states depend only on the variables $q_n$.\\
It is
important to see that the scalar constraint {\em forced} us to let the
reduced configuration space (with respect to the scalar constraint) not only
depend on the original configuration variables (before reducing) but also on
the original momenta ! This mixing of original polarizations is typical for
genuinely quadratic constraints with respect to the momenta and can be viewed
as a source of difficulty in the Dirac approach to select physical states.\\
The Dirac observables $q_n,p^n$ are both real and have both infinite 
range according to the agreement that we have made above.
The unique inner product that turns $q_n,p^n$ into self-adjoint operators
(for a specific choice for n)
would be the one corresponding to the Lebesgue measure $dq_n$. Since we are
dealing with functions that are defined on the infinite direct product space
$\times_{n=0}^\infty R$ we have the problem that the formal "infinite product
Lebesgue measure" $\bigwedge_{n=0}^\infty dq_n$ which would turn all the 
$q_n,p^n$
into self-adjoint operators is ill-defined unless we integrate only 
cylindrical functions. A solution is to go to the Bargman-Fock representation
and to work with the observables $a_n:=q_n+ip^n\mbox{ and }\bar{a}_n$. As is
well known, the  
measure $\mu$ that implements the reality conditions $(\hat{a}_n)^\dagger
=\hat{\bar{a}}_n$ is Gaussian and therefore we can actually integrate 
holomorphic functions on  $X:=\times_{n=0}^\infty C$ that
depend on an infinite number of variables. 
We therefore end up with 
$L_2(X,\mu)$, the usual Bargman-Fock Hilbert space.

\subsection{Dynamics of the model - deparametrization for
an infinite number of degrees of freedom}

What we have obtained up to now is the complete set of Dirac observables
which are time-independent. However, their interpretation is still
missing. On the other hand, interpretation is usually straightforward
for non-gauge invariant quantities. In order to arrive at an interpretation
there are two possible strategies available : either one expresses an object
that is $O(2)\times Diff(\Sigma)$ invariant in terms of Dirac observables
plus an additional gauge function t which labels the choice of foliation
of spacetime and quantizes only the Dirac observables (\cite{10}) or one uses
the framework of time-dependent Dirac observables (deparametrization,
\cite{6}).\\
Let us try to apply the latter one.\\
In the sequel we will only deal with the 'physically relevant' sector I.\\
We are asked to cast the scalar constraint into a form that allows to apply
an extension to an infinite number of degrees of freedom of the framework of
deparametrization. As is proved in (\cite{9}) the only change is
that the time variable becomes 'many-fingered' (Tomonaga-Schwinger
extension).\\
In particular we are interested in the
time-development of the quantities $E^x,E^y,E^z$ which determine the metric.
\\
The starting point will be equation (3.19). We take the square root and
restrict to the positive branch. Then the scalar constraint adopts the
familiar form
\beq P^0(z)-H(Q_1,P^1;Q_2,P^2)(z):=P^0(z)-\sqrt{[P^1(z)]^2+[P^2(z)]^2}
\eeq
which is already deparametrized, although at each point x of the hypersurface
separately. In (4.2) we have defined an infinite number of 'Hamiltonians',
$H(x)$, that is, an 'energy density'. Note that, since H involves only
momenta, there are no ordering problems.\\
Following this framework, we construct the time-evolution operator
\beq \hat{U}_t:=\exp(-i\int_\Sigma dz(Q_0-t)\hat{H}) \eeq
and define any operator $\hat{O}$ (to be factor-ordered appropriately) on the 
set
of physical states $\Psi=\hat{U}\psi$ (with respect to the scalar constraint)
by \beq
(\hat{O}_t\Psi)[Q_0,Q_1,Q_2]:=\hat{U}_t\hat{O}_{t=Q_0}\psi[Q^1,Q^2]\;.\eeq
Now, in contrast to toy models for cosmology, physical states should also
be diffeomorphism invariant. We will therefore take the operator to be
diffeomorphism-invariant, but not necessarily invariant under the scalar
constraint.\\
We want to apply the above framework for example to the operator for the
three-volume whose classical counterpart is given by
\beq V:=\int_\Sigma dz\sqrt{\det{q}}=\int_\Sigma dz\sqrt{E^x E^y E^z}\;.\eeq
We will proceed as follows : We start from (4.2)
and proceed to the Dirac quantization of the model. As for the scalar
constraint, we have no operator-ordering problems and the complete solution
of
\beq (-[\hat{P}^0]^2+[\hat{P}^1]^2+[\hat{P}^2]^2)\Psi[Q_0,Q_1,Q_2]=0 \eeq
is formally given by a functional integral (positive energy solutions)
\begin{eqnarray}
\Psi[Q_0,Q_1,Q_2] & = &\int_{R^2}[d^2k]\phi[k^1,k^2]
\exp(-i\int_\Sigma dz [\sqrt{(k^1)^2+(k^2)^2}Q_0-k^1 Q_1-k^2 Q_2])
\nonumber\\
& = & \exp(-i\int_\Sigma dzQ_0\sqrt{(\hat{P}^1)^2+(\hat{P}^2)^2})\psi[Q_1,Q_2]
\end{eqnarray}
where $[d^2k]\phi[k^1,k^2]$ is the formal expression for the limes
of a cylindrical measure (formally, $[d^2k]$ is an infinite Lebesgue
measure and $\phi$ is the Fourier transform of $\Psi$). The integral
is defined through a lattice evaluation and taking the limes that the
lattice spacing goes to zero at the end (the lattice
is finite since $\Sigma$ is compact).\\
Since $\hat{U}_t$ commutes with the vector constraint (ordered with the
momenta to the right), diffeomorphism invariance is obeyed if and only
if for an infinitesimal density $\xi$ of weight -1 holds
\beq \psi[Q_1+(Q_1)'\xi,Q_2+(Q_2)'\xi+\xi]=\psi[Q_1,Q_2] \;.\eeq
Inserting, the integrand of the path integral becomes after doing an
integration by parts in the exponential (up to first order in $\xi$)
\[ \phi[k^1,k^2]\exp(i\int_\Sigma dx[(k^1-(\xi k^1)')Q_1+(k^2-(\xi k^2)')Q_2
+\xi' k^2]) \;.\]
We change to new variables $K^i:=k^i-(\xi k^i)',\;i=1,2$ in the functional
integral and obtain (up to first order in $\xi$) the condition
\beq \phi[K_1+(K^i\xi)',K^2+(K^2\xi)']\exp(i\int_\Sigma dz \xi' K^2)
=\phi[K^1,K^2] \eeq
where we have used the diffeomorphism invariance of the Lebesgue measure
(the Jacobian is 1 to first order in $\xi$ : denote by h the spacing of the
lattice, $k=k^1\mbox{ or }k^2,\;k_n:=k(nh)$, then the Jacobian is
\begin{eqnarray}
& &  \frac{\partial k_n}{\partial K_m}=\frac{\partial[K_n-1/(2h)
((\xi K)_{n+1}-(\xi K)_{n-1})]}{\partial K_m}\nonumber\\
& = & \delta_{n,m}-\frac{\xi_m}{2h}(\delta_{n,m-1}-\delta_{n,m+1})
=:J_{nm}=:(1+A(\xi))_{n,m}
\end{eqnarray}
where the matrix A is trace-free and of first order in $\xi$, so
$\det(J)=1+\mbox{tr}(A)+..=1+O(\xi^2)$).\\
We now split $\phi[k^1,k^2]=:\exp(i\int_\Sigma dz f(k^2))\tilde{\psi}[k^1,
k^2]$
and seek to determine the function f according to the requirement that
\[ \int_\Sigma dx[f(K^2+(\xi K^2)')+\xi'K^2]=\int_\Sigma dx f(K^2) \]
up to first order in $\xi$. We expand f and perform integrations by parts
until only $\xi$ appears derivated with respect to z. Then a sufficient
condition for the last equation to hold is that
\[ k^2(1+\frac{\partial f(k^2)}{\partial k^2})-f=0\; . \]
The solution of this ordinary first order differential equation is given by
\[ f(k^2)=[c-\ln(k^2)]k^2 \]
where c is a real integration constant.\\
This means that $\tilde{\psi}$
is diffeomorphism invariant when $k^1,k^2$ transform as densities of weight
one. Again we are therefore lead to the functionals
\beq \tilde{\psi}[k^1,k^2]=f(\{q^n\}),\mbox{ where } q^n=\int_\Sigma d\mu
(d^n(k^2/k^1))^2 \eeq
with $d\mu=dz k^1,\;d=1/k^1\partial$.\\
Diffeomorphism invariance up to first order implies diffeomorphism invariance
with respect to the component of the identity of the diffeomorphism group.\\
We are now in the position to define the operator $\hat{V}$.\\
By making use of the definitions we can express (4.5) in terms of
$Q_\mu,P^\mu$
\begin{eqnarray}
E^{x/y} & = & \frac{p^{x/y}}{A_{x/y}(p^y+p^y)}\sqrt{(p^x+p^y)^2-(\Pi_\gamma')^2}
\mbox{ hence}\nonumber\\
\sqrt{\det(q)} & = & \exp(-Q^0)\frac{\sqrt{(P^0-P^2)^2-(P^1)^2}}{2(P^0-P^2)}
\nonumber\\
& & \sqrt{4(P^0-P^2)^2-[(\exp(-(Q_2+2Q_0))P^2)']^2}
\end{eqnarray}
and we face severe factor ordering problems when
promoting the integral over this function to quantum theory.\\
The action of this operator on physical states is given by (we assume that
we order all momenta to the right hand side in the formal power expansions
of the non-analytical terms in (4.12))
\begin{eqnarray}
\hat{V}_t\Psi[Q_0,Q_1,Q_2]&=&\hat{U}_t\int[d^2k]\exp(i\int_\Sigma
dz[ Q_1 k^1+Q_2 k^2+[c-\ln(k^2)]k^2])\tilde{\psi}[k^1,k^2]\times\nonumber\\
& & \int_\Sigma dx\exp(-t)\frac{\sqrt{(k^2)^2-(k^1)^2}}{(-2)k^2}
\sqrt{4(k^2)^2-[(\exp(-(Q_2+2t))k^2)']^2}\nonumber\\
& &
\end{eqnarray}
which, when acting with the evolution operator (which, by the way, is unitary
for $Q_0-t=$const. with respect to the inner product (4.1) : since the
integrated Hamiltonian commutes with both constraints, it is physical, that
is, well-defined on ${\cal H}_{phys}$, and
since it is classically real it promotes to a self-adjoint operator by a
suitable choice of ordering) results in a horrible object
which we cannot work out explicitly.\\
The point of this subsection was to show how deparametrization works in
principle for field theories and how one can arrive at an interpretation. It
also shows that most of the evolving
operators which one would intuitively think of as observables are probably
not well-defined on the physical Hilbert space (here in the connection
representation). \\

\section{Loop variables}

The concluding section of this paper is designated to the issue of using
(traces of) Wilson-loop observables for the quantization of or even for the
classical treatment of gravity. Because our model system is completely
integrable one can ask and {\em answer} all kinds of questions that have
been raised in the literature.\\
In particular we ask if the operators that are usually defined for the loop
representation in the literature (\cite{7}) are well defined at all, that is,
if they make sense when applied
to physical states. Of course, as they stand they will not leave the
physical Hilbert space invariant and thus are not even expected to have
a well-defined action on physical states. However, it will be interesting
to see how the Dirac observables look in terms of loop-coordinates, in
particular the constraints, what their Poisson structure is and how
one could define a 'loop-transform' into the loop representation (\cite{7}),
from states that have been defined for the connection representation.\\
Also, one could construct loop-operators as certain 'time-dependent' Dirac
observables.\\
We will deal here only with the definition of appropriate loop variables
in the connection representation.\\
\\
We begin by recalling that one defines the following 'basic' loop
variables when dealing with the loop representation (starting from the 
connection representation) :\\
The $T^0$ variables are defined to be traces of the holonomies (parallel
transport operators)
\beq U_\sigma[A](x):={\cal P}\exp(\oint_\sigma A) \eeq
around closed curves (loops $\sigma$) with base point x in the fundamental
representation of SU(2) (the Ashtekar variables are complex-valued whence
one they lie in the Lie algebra of complexified SU(2,C)=SL(2,C); 
although the gauge group of gravity is really SU(2), the loop variables 
are even invariant under SL(2,C))
\beq T_\sigma[A]:=\frac{1}{2}\mbox{tr}[U_\sigma[A]] \eeq
where $\cal P$ denotes path-ordering and $A:=A_a^idx^a\tau_i,\;\tau_i$ a
basis in the Lie algebra su(2). We will choose $\tau_i:=-i/2\sigma_i$, the
latter being the usual Pauli matrices. Note that the trace of the holonomy
is independent of the base-point on the given loop.\\
The $T^1$-variables are defined to be
\beq T^a_\sigma[A,E](x):=\mbox{tr}[E^a(x)U_\sigma[A](x)]\;. \eeq
As one can show, they form a closed Poisson-algebra and can thus serve as
a starting point for the group-theoretical quantization scheme. Let us
explore what happens with these variables for our model.\\
The (pull-back under a section of the principal SU(2)-bundle of the)
connection reduces to
\beq A(x,y,z)=A_x(\cos(\alpha)\tau_1+\sin(\alpha)\tau_2)dx
             +A_y(-\sin(\alpha)\tau_1+\cos(\alpha)\tau_2)dy
             +A_z\tau_3 dz \eeq
and the gauge transformations generated by the Gauss-constraint for our
reduced configuration variables are those under the U(1)
subgroup of SU(2) generated by $\tau_3$ (even its complexification) :
\beq -dS S^{-1}+S A S^{-1}\mbox{ where }S=\exp(\Lambda\tau_3) \eeq
can be written as in (5.4) only that $\alpha,A_z$ are replaced by
$\alpha+\Lambda,A_z-\Lambda'$. Under a gauge transformation, the holonomy
and the triad $E=E^a_i\tau_i\partial_a$ transform covariantly
\beq E(x)\rightarrow S(x)E(x)S^{-1}(x)\mbox{ and } U_{\sigma(x)}\rightarrow
S(x)U_\sigma(x)S^{-1}(x) \;.\eeq
whence the $T^0,T^1$ variables live on the Gauss-reduced phase space.\\
It is convenient to eliminate the angle $\alpha$ by doing an U(1,C) rotation
with gauge parameter $\Lambda=-\alpha$ (note that $\alpha$ is complex so
this transformation is really an element of U(1,C)). Then the connection and
the triad adopt the reduced structure
\begin{eqnarray}
A(x,y,z) &=& A_x(z)\tau_1dx+A_y\tau_2(z)dy+\gamma(z)\tau_3dz \\
E(x,y,z) &=& [\Pi^x(z)\partial_x-E^y(z)\sin(\alpha-\beta)\partial_y]\tau_1
+[\Pi^y(z)\partial_y-E^x(z)\sin(\alpha-\beta)\partial_x]\tau_2\nonumber\\
& & +\Pi^\gamma(z)\tau_3\partial_z \;.
\end{eqnarray}
Note that both quantities do not have any x,y dependence !\\
We will now start to find suitable loops to capture the full U(1) invariant
information needed in order to actually coordinatize the Gauss-reduced phase
space. Three loops are obvious : they are those lying entirely in one of the
three $S^1$-factors of the three-torus. We will call them $\sigma(x),
\sigma(y)\mbox{ and }\sigma(z)$ respectively. A short calculation yields
for the corresponding $T^0,T^1$'s (the range of the three coordinates
x,y and z is taken to be the interval $[0,1]$ with endpoints identified) :\\
\begin{eqnarray}
T_{\sigma(x)}(z)& = &\cos(\frac{1}{2}A_x(z)) \\
T_{\sigma(y)}(z)& = &\cos(\frac{1}{2}A_y(z)) \\
T_{\sigma(z)}   & = &\cos(\frac{1}{2}\int_0^1 dz\gamma(z)) \\
T^x_{\sigma(x)}(z) & = & -\Pi^x(z)\sin(\frac{1}{2}A_x(z))\\
T^y_{\sigma(y)}(z) & = & -\Pi^y(z)\sin(\frac{1}{2}A_y(z))\\
T^z_{\sigma(z)}(z) & = & -\Pi^y(z)\sin(\frac{1}{2}\int_0^1 dz\gamma(z))\\
T^x_{\sigma(y)}(z) & = & \frac{(\Pi_x\Pi^\gamma)(z)}{(A_x\Pi^x+A_y\Pi^y)(z)}
                         \sin(\frac{1}{2}A_y(z)) \\
T^y_{\sigma(x)}(z) & = & \frac{(\Pi_y\Pi^\gamma)(z)}{(A_x\Pi^x+A_y\Pi^y)(z)}
                         \sin(\frac{1}{2}A_x(z)) \\
& & T^x_{\sigma(z)}(z)=T^y_{\sigma(z)}(z)=T^z_{\sigma(x)}(z) \;.
=T^z_{\sigma(y)}(z)=0
\end{eqnarray}
Note that from the four quantities $T_{\sigma(\mu)},
T^\mu_{\sigma(\mu)}(z),\;
\mu\in\{x,y\}$ we can already extract the two gauge invariant canonical
pairs $A_\mu,\Pi^\mu$ {\em locally} while the two variables
$T^z_{\sigma(z)}(z),T_{\sigma(z)}$ allow to recover $\Pi^\gamma$
{\em locally} but only global information about $\gamma$ through the 
integrated quantity $\int_\Sigma dz\gamma(z)$.
This has two consequences :\\
1) The information contained in the two 'off-diagonal' variables
$T^\mu_{\sigma(\nu)}(z),\;\mu\not=\nu$ is already redundant and so we can
neglect them,\\
2) We need a further $T^0$ variable in order to get hand on $\gamma(z)$.\\
Before doing that we show that the variables defined above form a closed
(classical) *-Poisson algebra (the Loop-variables for the Ashtekar phase
space are {\em not} closed under complex conjugation at least it is not
obvious how to express $\bar{T}_\alpha=\frac{1}{2}\mbox{tr}({\cal P}
\exp(\int_\alpha[2\Gamma-A]))$
in terms of $T^0,T^1\mbox{ where }\Gamma$ is the spin-connection). \\
We have proved in the sections before that $A_x,A_y,\gamma$ are purely
imaginary while $\Pi^x,\Pi^y,\Pi^z$ are purely real. This implies the
*-relations
\beq  \overline{T_{\sigma(a)}(z)}=T_{\sigma(a)}(z)\mbox{ and }
\overline{T^a_{\sigma(a)}(z)}=-T^a_{\sigma(a)}(z)
\eeq
i.e. the $T^0$'s are purely real, the $T^1$'s purely imaginary, and also that
the range of the $T^0$ are the real numbers larger than or equal to one while
the $T^1$'s have the whole imaginary axis as their range. This opens the
chance for a cotangent topology of the associated phase space and therefore
a group theoretical quantization scheme.\\
Another consequence is that the variables $T_{\sigma(\mu)}$ are not
{\em quite} sufficient to recover the connection up to gauge
transformations : we have the degeneracy
\beq T_{\sigma(\mu)}[A]=T_{\sigma(\mu)}[-A] \eeq
which means that these loop variables are not separating on the moduli space
${\cal A}/{\cal G}$ of SL(2,C) connections modulo gauge transformations for
our model when restricting ourselves` to the above defined class of loops.
This is in analogy to the known degeneracies (\cite{13}) of
these loop variables as coordinates for ${\cal A}/{\cal G}$\footnote{This
result implies, for example, that any topology on ${\cal A}/{\cal G}$ which
is based on the use of the loop variables (traces of the holonomy) displays
${\cal A}/{\cal G}$ as a non-Hausdorff space (\cite{14}) !}.
 \\
Next we compute the Poisson-structure which turns out to be quite similar
to the one obtained in the first reference of (\cite{8}) when restricting
ourselves to the\\
 z-direction :
\begin{eqnarray}
\{T_{\sigma(a)}(z),T_{\sigma(b)}(z')\}& = & 0\\
\{T_{\sigma(\mu)}(z),T^\nu_{\sigma(\nu)}(z')\}& = &
\frac{1}{2}\delta_{\mu\nu}\delta(z,z')\sin^2(\frac{1}{2}A_\mu(z))
=\frac{1}{2}\delta_{\mu\nu}\delta(z,z')(1-[T_{\sigma(\mu)}(z)]^2)\nonumber\\
 & & \\
\{T_{\sigma(\mu)}(z),T^z_{\sigma(z)}(z')\}& = &
\frac{1}{2}\sin^2(\frac{1}{2}\int_0^1\gamma(z)dz)
=\frac{1}{2}(1-[T_{\sigma(z)}]^2)\\
\{T^\mu_{\sigma(\mu)}(z),T^\nu_{\sigma(\nu)}(z')\}
& = &\{T^\mu_{\sigma(\mu)}(z),T^z_{\sigma(z)}(z')\}=0\\
\{T^z_{\sigma(z)}(z),T^z_{\sigma(z)}(z')\} & = &
\frac{1}{2}[T^z_{\sigma(z)}(z)-T^z_{\sigma(z)}(z')]T_{\sigma(z)}
\end{eqnarray}
and we could even have written the right hand side of all the equations as
linear combinations of single $T^0\mbox{'s or },T^1$'s with
distribution-valued coefficients at the price of extending the so defined
loop algebra to all such loops with any winding number (with respect to an
orientation on the three torus) by making use of the Mandelstam identities
\beq T_\alpha[A] T_\beta[A]=\frac{1}{2}[T_{\alpha\circ\beta}
+T_{\alpha\circ\beta^{-1}}] \eeq
(\cite{6}) where $\circ$ is the product in the loop group (with respect to
a given base point). This Mandelstam identity becomes here just the
trigonometrical identity
\beq \cos(\alpha)\cos(\beta)=\frac{1}{2}[\cos(\alpha+\beta)
+\cos(\alpha-\beta)] \eeq
but we refrain from introducing the winding number because we will have to
deal with higher order polynomials of loop variables anyway.\\
Note incidentally that (5.21) implies that the the conjugate variable
to $T^\mu_{\sigma(\mu)}$ is given by $2\mbox{arth}(T_{\sigma(\mu)})$ !\\
Remark :\\
It is clear that the Hamilton-Jacobi functional $S:=cT_{\sigma(z)}$ solves
all constraints for the degenerate sector for any constant c. Thus 
it turns out that the loop variables serve as 'good' coordinates only for 
(part of) the degenerate sector as for the full theory (\cite{21}).\\
\\
We now have to extract $\gamma(z)$. One way would be to make use of the
so-called area-derivative (\cite{15}), whose relevant components
for the already defined loops become (the area-derivative in the (ab)-plane
is denoted $\delta_{ab}$) at the point with coordinate z (x,y are irrelevant)
\begin{eqnarray}
\delta_{xy}T_{\sigma(z)} & = & -\frac{1}{2}A_x(z)A_y(z)
\sin(\frac{1}{2}\int_0^1dz\gamma(z)) \\
\delta_{yz}T_{\sigma(x)} & = & -\frac{1}{2}A_y(z)\gamma(z)
\sin(\frac{1}{2}A_x(z)) \\
\delta_{zx}T_{\sigma(y)} & = & -\frac{1}{2}\gamma(z)A_x(z)
\sin(\frac{1}{2}A_y(z))
\end{eqnarray}
but since the area-derivative is a certain limit of a difference between
one of the above loop variables and an extended one (the area derivative in
the (ab)-plane with respect to a given loop at one of its points is obtained
by attaching an infinitesimal loop lying in the (ab)-plane to the given loop
at the point of interest, computing the trace of the holonomy
for this modified loop and the original one, taking the difference and
dividing by the area element) we actually leave the above algebra and look
at an infinite number of new loops. We would like to have only as many
loop variables as reduced canonical variables. Accordingly, we are looking
for a better suited object.\\
After some trial, one of the most simple objects seems to be an eyeglass
loop (\cite{7}) obtained by drawing two loops with equal orientation at
$z_1\mbox{ and }z_2$ in x or y direction connected by a line in z-direction
back and forth between $z_1\mbox{ and }z_2$.\\
We will denote this loop by $\sigma(xz)\mbox{ or }\sigma(yz)$. The
computation of the associated $T^0$ reveals
\beq T_{\sigma(xz)}(z_1,z_2)=\cos(\frac{1}{2}A_x(z_1))
\cos(\frac{1}{2}A_x(z_2))-\cos(\int_{z_1}^{z_2}dz\gamma(z))
\sin(\frac{1}{2}A_x(z_1))\sin(\frac{1}{2}A_x(z_2)) \eeq
and thus allows to extract $\gamma(z)$ by taking the derivative with respect
to, say, $z_2=z$.\\
Fix a value $z_0$ on the z-circle and write
\beq T_{\sigma(xz)}(z):=T_{\sigma(xz)}(z_0,z) \;.\eeq
We observe that
\beq T_{\sigma(xz)}(z_0+1)=[T_{\sigma(x)}(z_0)]^2-(2[T_{\sigma(z)}]^2-1)
(1-[T_{\sigma(x)}(z_0)]^2) \eeq
whence
\beq T_{\sigma(z)}=\sqrt{\frac{1-T_{\sigma(xz)}(z_0)}{2(1-
[T_{\sigma(x)}(z_0)]^2)}}\eeq
so that the information contained in $T_{\sigma(z)}$ is also contained in
$T_{\sigma(xz)}(z)$ (if one uses also $T_{\sigma(\mu)}(z)$) (note that
even its sign can be determined from (5.23) as the $T_{\sigma(a)}(z)$ are
all positive). Therefore we will consider $T_{\sigma(z)}$ as redundant
in the following although we will use it as an abbreviation for (5.33). Also
the analogon $T_{\sigma(yz)}\mbox{ of }T_{\sigma(xz)}$ is not needed.\\
We now compute the extended loop algebra : the only non-vanishing brackets
of our new element with the rest are (up to a sign)
\begin{eqnarray}
& & \{T_{\sigma(xz)}(z),T^z_{\sigma(z)}(z')\}=-\chi_{[z_0,z]}(z')
\sin(\int_{z_0}^{z}\gamma(t)dt)\sin(\frac{1}{2}\int_0^1dt\gamma(t))
\nonumber\\
& & \sin(\frac{1}{2}A_x(z_0))\sin(\frac{1}{2}A_x(z))=-\chi_{[z_0,z]}(z')
\times \nonumber\\
& & \times\sqrt{(1-[T_{\sigma(z)}]^2)[(1-[T_{\sigma(x)}
(z_0)]^2)(1-[T_{\sigma(x)}(z)]^2)-(T_{\sigma(xz)}(z)-T_{\sigma(x)}
(z_0)T_{\sigma(x)}(z))^2]}\nonumber\\
& & \\
& & \{T_{\sigma(xz)}(z),T^x_{\sigma(x)}(z')\}=\frac{1}{2}(\delta(z',z_0)
[-T^x_{\sigma(x)}(z)+T^x_{\sigma(x)}(z_0)T_{\sigma(xz)}(z)]\nonumber\\
& & +\delta(z',z_1)[-T^x_{\sigma(x)}(z_0)+T^x_{\sigma(x)}(z)T_{\sigma(xz)}(z)])
\end{eqnarray}
Unfortunately, the right hand side becomes quite complicated and it would seem
to be better to work with an overcomplete set and apply the algebraic
quantization programme (\cite{17}). In our case that would boil
down to the problem to find loop-variables that depend only on the
configuration variables, contain $\sin(\frac{1}{2}A_\mu(z)),\sin(\int_0^1
dz\gamma(z)/2)$ and yield brackets that have analytical right hand sides
for the corresponding Poisson-brackets, but that turns out not to be
straightforward without regarding the whole set of loops.\\
We now come to set up a bijection (up to signs) between the Loop-variables
defined so far and the canonical variables determined in section 3.\\
We have already defined the T's in terms of $A_\mu,\Pi^\mu;\gamma,
\Pi^\gamma$. Here is the local\\
 inversion :
\begin{eqnarray}
A_\mu(z) & = & 2\mbox{arcos}(T_{\sigma(\mu)}(z)) \\
\Pi^\mu(z) & = & -\frac{T^\mu_{\sigma(\mu)}(z)}{\sqrt{1-[T_{\sigma
(\mu)}(z)]^2}} \\
\gamma(z) & = & [\mbox{arcos}(\frac{T_{\sigma(x)}(z_0)T_{\sigma(x)}(z)
-T_{\sigma(xz)}(z)}{\sqrt{(1-[T_{\sigma(x)}(z)]^2)
(1-[T_{\sigma(x)}(z_0)]^2)}})]' \\
\Pi^\gamma(z) & = & -\frac{T^z_{\sigma(z)}(z)}{1-\frac{\sqrt{1-T_{\sigma(xz)}
(z_0)}}{2(1-[T_{\sigma(x)}(z_0)]^2)}}
\end{eqnarray}
where the prime denotes derivation with respect to the z-argument only. The
z-derivation considered as a difference of Loop-variables does not lead out
of the class of variables that we have defined and so we are fine.\\
(5.36)-(5.39) defines a (local) {\em point} transformation which is always
a local
symplectomorphism. Accordingly, when plugging the above formulae into the
symplectic potential, we are actually able to determine the momenta
conjugate to the three variables $T_{\sigma(\mu)},T_{\sigma(xz)}$ !\\
After some tedious algebraic manipulations, an integration by parts,
using that
$\int_{S^1}\gamma(z)dz$ written in terms of $T^0$'s is a spatial constant
and making use of the fact that quantities frozen at $z_0$ are also
spatial constants we derive the momenta conjugate to\\ $T_{\sigma(x)}(z),
T_{\sigma(y)}(z),T_{\sigma(xz)}(z)$, the basic loop variables that we are
using, respectively as
\begin{eqnarray}
& & \frac{[T^z_{\sigma(z)}(z)]'[T_{\sigma(x)}(z_0)-T_{\sigma(x)}(z)
T_{\sigma(xz)}(z)]} {\sqrt{(1-[T_{\sigma(z)}]^2)[(1-[T_{\sigma(x)}(z_0)]^2)
(1-[T_{\sigma(x)}(z)]^2)-(T_{\sigma(xz)}(z)-T_{\sigma(x)}(z_0)
T_{\sigma(x)}(z))^2]}}\times \nonumber\\
& & \times\frac{1}{1-(T_{\sigma(x)}(z))^2}
-\frac{T^x_{\sigma(x)}(z)}{1-[T_{\sigma(x)}(z)]^2},  \\
& & -\frac{T^y_{\sigma(y)}(z)}{1-[T_{\sigma(y)}(z)]^2}, \\
& & -\frac{[T^z_{\sigma(z)}(z)]'}{\sqrt{(1-[T_{\sigma(z)}]^2)
[(1-[T_{\sigma(x)}(z_0)]^2)(1-[T_{\sigma(x)}(z)]^2)
-(T_{\sigma(xz)}(z)-T_{\sigma(x)}(z_0)T_{\sigma(x)}(z))^2]}}\nonumber\\
& &
\end{eqnarray}
and the momentum conjugate to $T_{\sigma(x)}(z_0)$ is the integral over
$S^1$ of expression (5.40) only that the roles of $T_{\sigma(x)}(z_0)
\mbox{ and }T_{\sigma(x)}(z)$ are interchanged. The reason for the appearance
of this global degree of freedom is that we fixed the point $z_0$ to be
an observable by hand. \\
Next we look at the constraints expressed in terms of loop-variables. From
their definition it is obvious that $T_{\sigma(\mu)}$ is a scalar, and
since (5.38) shows that the density $\gamma$ is a spatial derivative of
a function of our three loop variables, we find that also $T_{\sigma(x,z)}$
must be a scalar (alternatively, this follows from its definition).
$T_{\sigma(x)}(z_0)$ is a constant and its conjugate, as an integral over
a density, is also diffeomorphism invariant.\\
Thus, necessarily the vector constraint must be
\beq (T_{\sigma(\mu)})'S^\mu+(T_{\sigma(xz)})'S \eeq
where we have denoted the corresponding conjugate momenta in (5.40), (5.41)
by $S^\mu,S$. Of course, it is also obvious from the definitions that
$T^\mu_{\sigma(\mu)}\mbox{ and }T^z_{\sigma(z)}$ are densities and a
scalar respectively.\\
It turns out to be more convenient to make direct use of the inversion
formulas (5.36)-(5.39) to obtain the explicit expressions :
\begin{eqnarray}
V & = & 2[\frac{T_{\sigma(x)}'T^x_{\sigma(x)}}{1-(T_{\sigma(x)})^2}
+\frac{T_{\sigma(y)}'T^x_{\sigma(y)}}{1-(T_{\sigma(y)})^2}\nonumber\\
& & +[\mbox{arcos}(\frac{T^0_{\sigma(x)}T_{\sigma(x)}
-T_{\sigma(xz)}}{\sqrt{(1-[T_{\sigma(x)}]^2)
(1-[T^0_{\sigma(x)}]^2)}})]'
\frac{(T^z_{\sigma(z)})'}{2-\frac{\sqrt{1-T^0_{\sigma(xz)}}}
{1-[T^0_{\sigma(x)}]^2}}] \\
C & = & 4[\frac{\mbox{arcos}(T_{\sigma(x)})T^x_{\sigma(x)}}{\sqrt{1-
[T_{\sigma(x)}]^2}}\frac{\mbox{arcos}(T_{\sigma(y)})T^y_{\sigma(y)}}{\sqrt
{1-[T_{\sigma(y)}]^2}} \nonumber\\
& & +[\frac{\mbox{arcos}(T_{\sigma(y)})T^y_{\sigma(y)}}{\sqrt
{1-[T_{\sigma(y)}]^2}}+\frac{\mbox{arcos}(T_{\sigma(x)})T^x_{\sigma(x)}}
{\sqrt{1-[T_{\sigma(x)}]^2}}]\times\nonumber\\
& & \times[\mbox{arcos}(\frac{T^0_{\sigma(x)}T_{\sigma(x)}(z)
-T_{\sigma(xz)}(z)}{\sqrt{(1-[T_{\sigma(x)}]^2)
(1-[T^0_{\sigma(x)}]^2)}})]'
\frac{T^z_{\sigma(z)}}{2-\frac{\sqrt{1-T^0_{\sigma(xz)}}}
{1-[T^0_{\sigma(x)}]^2}}]
\end{eqnarray}
where all expressions are to be evaluated at z except for those with a 0
superscript which are to be evaluated at $z_0$.\\
The result looks much more complicated than when using the variables of the 
previous section.
From the point of view of the loop representation one
would have to take equations (5.44) and (5.45) and apply it (after a
choice of ordering) to a loop state (\cite{7}) according to the Rovelli
Smolin loop representation of $T^0,T^1$. In order to do that one would
have to expand the trigonometrical formulas involved in both constraints and
one would obtain an expression involving an infinite sum of loop states. This 
is a very discouraging result, the more as it seems that the expressions 
(5.44) and (5.45) are non-analytical in nature so that it is unclear how to 
define the action of both constraints on loop states in the 
loop representation.\\
It is clear how to obtain the observables of the theory : simply substitute
the loop variables via (5.36)-(5.39) into the expressions for the
observables given in section 3. These formulas are too long and it is not
worthwhile as to display them here.\\

\section{Conclusions}

Let us briefly summarize what we have learnt by studying this model
system :\\
First of all we worked out a number of ideas and methods to reduce a
spatially diffeomorphism invariant field theory which might be useful
to keep in mind for future work. In particular, our methods do not
suffer from the problem of the existence of a bijection between a frame on
the hypersurface $\Sigma$ and a collection of scalar field to solve the
diffeomorphism constraint as it occurs in the second reference of
\cite{4}).

The symplectic reduction of the diffeomorphism constraint has expectedly
lead to nonlocal 'already smeared' Dirac observables in a natural way.
It was not necessary to make use of a complete system of mode functions
on the hypersurface (exploiting the fact that the Hilbert space of square
integrable functions on $\Sigma$ is separable) to write the true degrees of
freedom in such a countable fashion.

The solution of the scalar constraint quadratic in the momenta reveals that
the Dirac observables will in general mix momentum and configuration
variables relative to the polarization of the phase space that one started
with.

The formalism of deparametrization applied to field theories shows that there
are even more serious problems than those that have been discussed for
finite-dimensional examples in \cite{4} at least on the technical side.

The fact that the loop-variables are so badly suited for our model
is quite discouraging and scratches at the hope that many
authors had when employing them for gravity, namely that they could
serve as the natural 'polarization' to solve quantum gravity
non perturbatively. Actually, the quantum solutions that have been found in
\cite{7} are solutions for the degenerate sector only. In our model, the 
variables $T_{\sigma(z)}\mbox{ and }T^z_{\sigma(z)}$ {\em are} simple 
expressions and the former is easily seen to be a Dirac observable on the
degenerate sector only. However, it is not the only solution of the constraints
(in \cite{7} it was speculated that the found solutions are complete).
The assumption arises that it will be a general feature that the solution of
the constraints expressed in terms of the loop variables will take a simple
form only on the degenerate sector.\\
On the other hand, one should never overestimate the results obtained from a 
model and it could in fact also happen that our conclusions are due to an
artifact of the reduction process.\\
Especially the fact that we are dealing with a restricted class of loops seems
to contribute to the complicate appearance of (5.44) and (5.45).\\
\\
\\
{\large Acknowledgements}\\
\\
Prof. A. Ashtekar has raised the problem of studying loop variables for
reduced models. This project was supported by the Graduiertenkolleg of the
Deutsche Forschungsgemeinschaft.

\begin{appendix}

\section{Calculation of the spin-connection}

We will derive here a more general result, namely how the spin-connection
looks like for the Gowdy models (\cite{8}).\\
The densitized triad for the Gowdy models
takes the form
\beq
(E^a_i)=
\left ( \begin{array}{*{2}{c@{\mbox{ }}}c}
        E^x_1 & E^y_1 & 0 \\
        E^x_2 & E^y_2 & 0 \\
        0 & 0 & E^z
        \end{array}
\right )
\mbox{ so that }\tilde{\tilde{q}}^{ab}=
\left ( \begin{array}{*{2}{c@{\mbox{ }}}c}
        E^x_I E^x_I & E^x_I E^y_I & 0 \\
        E^x_I E^y_I & E^y_I E^y_I & 0 \\
        0 & 0 & (E^z)^2
        \end{array}
\right )
\eeq
is the expression for the twice densitized inverse metric $E^a_i E^b_i$.
We conclude that
\beq [\det(q)]^2=[\det(E^a_i)]^2=(E^z\det(E^\mu_I))^2=:(E^z E)^2 \eeq
and can invert expression (A.1) to obtain the metric tensor
\beq
(q_{ab})=\frac{E^z}{E}
\left ( \begin{array}{*{2}{c@{\mbox{ }}}c}
        E^y_I E^y_I & -E^x_I E^y_I & 0 \\
        -E^x_I E^y_I & E^x_I E^x_I & 0 \\
        0 & 0 &(\frac{E}{E^z})^2
        \end{array}
\right ) \;.
\eeq
From this we infer the expressions for the triad and its inverse, that is
\beq e^a_i=\frac{E^a_i}{\sqrt{\det(q)}}\mbox{ and }e_a^i=q_{ab}e^b_i \eeq
namely
\beq
(e^a_i)=
\frac{1}{\sqrt{E^z E}}
\left ( \begin{array}{*{2}{c@{\mbox{ }}}c}
        E^x_1 & E^y_1 & 0 \\
        E^x_2 & E^y_2 & 0 \\
        0 & 0 & E^z
        \end{array}
\right )
\mbox{ and }(e_a^i)=
\sqrt{\frac{E^z}{E}}
\left ( \begin{array}{*{2}{c@{\mbox{ }}}c}
        E^y_2 & -E^x_2 & 0 \\
        -E^y_1 & E^x_1 & 0 \\
        0 & 0 & \frac{E}{E^z}
        \end{array}
\right )
\eeq
which we have to insert into the formula for the spin-connection 
\beq
\Gamma_a^i=-\frac{1}{2}\epsilon^{ijk}e^b_k(2 e_{[a,b]}^j+e^c_j e_a^l
e_{c,b}^l) \; .\eeq
Note that $e_{a,b}^i=(e_a^i)'\delta_{bz}$ in the course of the calculation
and that $e_z^i e^z_j\epsilon^{ijk}=e_z^i (e^z_j)'\epsilon^{ijk}=e_z^i
e^\mu_i=0$ etc. due to the symmetry properties of the triads so that we can
simplify (A.6) without reference to the explicit form of the triads to
\beq
\Gamma_a^i=-\frac{1}{2}\epsilon^{ijk}[e^z_k((e_a^j)'+e^c_j e_a^l
(e_c^l)')+\delta_{a z}e^b_j (e_b^k)'] \; .\eeq
More explicitly
\begin{eqnarray}
\Gamma_x^i & = & \frac{1}{2}\epsilon^{ijk}e^z_j((e_x^k)'+e^x_k e_x^l
(e_x^l)'+e^y_k e_x^l(e_y^l)' \nonumber\\
\Gamma_y^i & = & \frac{1}{2}\epsilon^{ijk}e^z_j((e_y^k)'+e^x_k e_y^l
(e_x^l)'+e^y_k e_y^l(e_y^l)' \nonumber\\
\Gamma_z^i & = & \frac{1}{2}\epsilon^{ijk}(e^x_j(e_x^k)'+e^y_j(e_y^k)') \;.
\end{eqnarray}
Now we have to go into details and have to make use of the expressions
(A.1). The calculation is rather tedious. The result is given by
\begin{eqnarray}
(\Gamma_x^i) & = & \frac{E^z}{2 E}(\frac{(E^z)'E^y_1}{E^z}+\frac{1}{E}
[(E^y_1)'E-(E^y_2)'E^x_I E^y_I+(E^x_2)' E^y_I E^y_I], \nonumber\\
& & \frac{(E^z)'E^y_2}{E^z}+\frac{1}{E}
[(E^y_2)'E+(E^y_1)'E^x_I E^y_I-(E^x_1)' E^y_I E^y_I],0) \nonumber\\
(\Gamma_y^i) & = & \frac{E^z}{2 E}(-\frac{(E^z)'E^x_1}{E^z}+\frac{1}{E}
[-(E^x_1)'E-(E^x_2)'E^x_I E^y_I+(E^y_2)' E^x_I E^x_I], \nonumber\\
& & -\frac{(E^z)'E^x_2}{E^z}+\frac{1}{E}
[-(E^x_2)'E+(E^x_1)'E^x_I E^y_I-(E^y_1)' E^x_I E^x_I],0) \nonumber\\
\Gamma_z^i & = & \frac{1}{2E}(0,0,E^x_I (E^y_I)'-E^y_I(E^x_I)')\;.
\end{eqnarray}
These are the corresponding real parts for the Ashtekar connection for the
Gowdy models. Let us restrict now (A.9) to our model, that is
\beq
(E^a_i)=
\left ( \begin{array}{*{2}{c@{\mbox{ }}}c}
        E^x\cos(\beta) & -E^y\sin(\beta) & 0  \\
        E^x\sin(\beta) & E^y\cos(\beta) & 0 \\
        0 & 0 & E^z
        \end{array}
\right ) \;.
\eeq
Then (A.9) simplifies tremendously to
\beq
(\Gamma_a^i)=
\left ( \begin{array}{*{2}{c@{\mbox{ }}}c}
        -\Gamma_x\sin(\beta) & \Gamma_y\cos(\beta) & 0  \\
        \Gamma_x\cos(\beta) & \Gamma_y\sin(\beta) & 0 \\
        0 & 0 & \Gamma_z
        \end{array}
\right )
\eeq
where
\beq
\Gamma_x=\frac{1}{2 E^y}(\frac{E^y E^z}{E^x})',\;\Gamma_y=-\frac{1}{2 E^x}
(\frac{E^x E^z}{E^y})',\;\Gamma_z=-\beta'\;.
\eeq

\end{appendix}

\end{document}